\documentclass[showpacs,showkeys,preprintnumbers,preprint]{revtex4}
\usepackage{amsmath}
\usepackage{graphicx}
\usepackage{dcolumn}
\usepackage{bm}
\usepackage{amssymb}
\usepackage{makeidx}
\usepackage{amsmath}
\usepackage{lscape}
\usepackage{epsfig}
\newcommand{\ri}{{\mathrm i}}
\newcommand{\p}{\partial}
\newcommand{\bea}{\begin{array}}
\newcommand{\eea}{\end{array}}

\newcommand{\la}{\label}

\begin{document}

\allowdisplaybreaks

  \title {Symmetries of the Schr\"odinger-Pauli equation for neutral
   particles}
\author{A.G. Nikitin}
\email{nikitin@imath.kiev.ua} \affiliation{ Institute of
Mathematics, National Academy of Sciences of Ukraine,\\ 3
Tereshchenkivs'ka Street, Kyiv, Ukraine, 01601}
 \date{\today}

\begin{abstract} With using the algebraic approach the Lie symmetries of Schr\"odinger equations with matrix potentials are classified. Thirty three inequivalent equations of such type together with the related symmetry groups are specified, the admissible equivalence relations are clearly indicated. In particular the Boyer results concerning kinematical invariance groups for arbitrary potentials (C. P. Boyer,  Helv. Phys. Acta, {\bf 47}, 450--605 (1974)) are clarified and corrected.
\end{abstract}

\maketitle

\section{Introduction\label{int}}
In addition to its dominant position in quantum physics,
Schr\"odinger equation is a very important mathematical subject
which stimulated the development (or even creation and development)
of fundamental research fields of the quin of sciences. A well known
example of such research fields is the inverse problem approach and
some special branches of functional analysis.

    Schr\"odinger equation is also a very important base for application
of various symmetry approaches to mathematical physics. Its symmetry
with respect to the eleven parameter continuous group in fact was
established long time ago by Sophus Lie. More exactly, Lie
discovered  the symmetries of the linear heat equation, but the SE
is nothing but its complex form. Then the Lie results were recovered
and developed in papers \cite{Nied}, \cite{And} and \cite{Boy}. It
was Niederer \cite{Nied} who had found the maximal invariance group
of the free Schr\"odinger equation. He  was the first who shows for
physicists  that in addition to the Galilei group, this group
includes also dilations and conformal transformations.

Symmetries of the one dimensional Schr\"odinger equation with
non-trivial potential were described in paper \cite{And}. Boyer
\cite{Boy} extends these results for  two- and three-dimensional
systems. The mentioned results occupy a place of honor in modern
physics. In fact they form the group-theoretical grounds of quantum
mechanics. The group classification presented in \cite{Boy} also is
a necessary step in investigations of  higher symmetries of
Schr\"odinger equation  which starts with the papers of Winternitz
with collaborators  \cite{Wint}, \cite{Wint1}, and in the search for
the coordinate systems which can be used for separation of variables
\cite{Mil}.

The higher symmetries of Schr\"odinger equations are nothing but
integrals of motion realized by differential operators whose  order
is higher than one. The complete description of 2d quantum
mechanical systems admitting second order integrals of motion was
presented in \cite{Wint}. And it had needed as much as twenty four
years to extend this result to the case of the 3d systems, see
papers \cite{evan} and \cite{evan2}.

The higher symmetries  give rise to such nice properties of SE as
superintegrability and supersymmetry, see surveys \cite{PW} and \cite{NNN}.
We will not discuss them here but mention that searching for such symmetries
is still a very popular business, and the modern trends in this field are
related to the third order and even arbitrary order integrals of motion
\cite{wintmar},  se also paper \cite{NikNik} where the determining equations
 for such symmetries were presented.

But there are important generalizations of SE which also play the
outstanding roles in theoretical physics. They  are
Schr\"odinger-Pauli (SP) equation and
  Schr\"odinger equation with position dependent mass (PDM equation).
Some of them  are superintegrable and supersymmetric and admit
various types of
  Lie symmetries. And it would be natural (and desirable) to extend the
  results concerning symmetry and superintegrability properties of the SE
  to the case of its mentioned generalizations.

It happens that  the group classification of Schr\"odinger equations
with PDM has been waited for a  long time. There is a lot of papers
devoted to PDM Schr\"odinger equations with particular symmetries,
see, e.g., \cite{11,rac,Koch,Cru}. Superintegrability aspects of
such equations (with trivial potentials) are discussed in
\cite{Fordy} and \cite{Fordy2}, see also the references cited
therein. But the complete group classification of these equations
appears rather recently in papers \cite{NZ}  and \cite{NZ2,NN} for
the stationary and time dependent equations correspondingly. In
paper \cite{N1} we start the systematic search for the higher order
symmetries in the PDM systems, but this program is not completed
yet.

Symmetries of the SP equations also are studied only partially. We
can mention  few papers devoted to its supersymmetries
\cite{N3,N4,kar}, extended supersymmetries, higher order symmetries
\cite{ N5,N6, N7} and Fock symmetries \cite{N8,N9,N10}. There are
also papers \cite{B} and \cite{M} where the relativistic aspects of
such symmetries are discussed.  However, in contrast with the
standard SE, we have no general group classification of these
equations. This circumstance  have to cause the blame for experts in
group analysis, taking into account the fundamental role played by
this  equation in quantum physics.

In the present paper we give the completed description of all
ineguivalent continuous symmetries which can be accepted by time
dependent SP equation. However, we restrict ourselves to SP equation
for neutral particles with spin, which have zero charge but non-zero
dipole momentum. A perfect example of such particle is the neutron.

We also present the corrected version of the Boyer classification of
continuous symmetries of the standard SE. First, these symmetries
form a subclass of symmetries of SP equation, and we need them to
formulate the  results of our research. Secondly, the Boyer
classification appears to be incomplete \cite{NN, Nuca}.

We believe that the physical community can pretend to a conveniently
presented and correct information on continuous symmetries which can
be admitted by the main equation of quantum mechanics, and use the
occasion to present the completed list of symmetries of both
Schr\"odinger and SP equations together with a clear definition of
the equivalence relations and notification of the corresponding
symmetry groups.

To solve the classification problem we use the so called algebraic
approach whose main idea is the a priori analysis of possible
symmetry algebras which are nothing but subalgebras of generic
invariance algebra of equations of interest. This approach makes it
possible to simplify the procedure of solving of the determining
equations and makes the classification results credible.

\section{Schr\"odinger-Pauli equation and its generalizations}

We will  consider SP equations of the following generic form
\begin{gather}\left(\ri\frac{\p}{\p t}-H\right)\psi(t,{\bf x})=0\la{se}\end{gather}
where $H$ is the Hamiltonian given by the following relation:
\begin{gather}H=-\frac12 \p_a\p_a
+V({\bf x})\la{H}\end{gather}
where
\begin{gather*}\p_a=\ri p_a=\frac{\p}{\p{x_a}},\quad {\bf x}=(x_1, x_2,x_3)\end{gather*}
and summation is imposed over the repeating indices $a$ over the values $a=1, 2, 3$. Moreover, in contrast with the standard SE, $V=V({\mathbf x})$ is not a scalar, but matrix potential which we expand via Pauli matrices:
\begin{gather}\la{mp}V= V_0+\sigma_aV^a.\end{gather}

If $V_0=0$ than equations (\ref{H}) and (\ref{mp}) define the standard SP equation for particles with trivial electric charge (in his case  $V_a$ should be  proportional to components of the external magnetic field). We reserve the possibility of nontrivial $V_0$ to be able to recover the case of the standard SE (in this case $V_a=0$) and the case of presence of a more generic scalar potential term not necessary the electric one.

We will search for symmetries of equations (\ref{se}) with respect
to continuous groups of transformations of dependent and independent variables. We will not apply the generic Lie approach whose perfect presentation can be found in \cite{olver} but restrict ourselves to its simplified version  adopted to {\it linear} equation (\ref{se}). Let us write the generator of the searched transformation group in the form
\begin{gather}\label{so}
    Q=\xi^0\partial _t+\xi^a\partial_a+\tilde\eta\equiv \xi^0\partial _t+
    \frac12\left(\xi^a\partial_a+\partial_a\xi^a\right)+\ri\eta,
\end{gather}
where $\tilde
\eta=\frac12\xi^a_a+\ri\eta,\ \ $ $\xi^0$, $\xi^a$ and $\eta$ are
functions of $t,{\mathbf x} $  and $\p_t=\frac{\p}{\p t}$. Moreover, $\eta$ is a $2\times 2$ matrix which, in analogy with (\ref{mp}), we represent in the following form:
\begin{gather}\la{eta1}\eta= \eta^0+\sigma_a\eta^a.\end{gather}
In contrary, $\xi^0$ and $\xi^a$ are scalar functions which can be treated as multipliers for the unit matrix.

Generator (\ref{so}) transforms solutions of equation (\ref{se}) into solutions if it satisfies the  following operator equation
\begin{gather}\la{ic}[Q,L]\equiv QL-LQ=\alpha L\end{gather}
where $L=\ri\p_t-H$ and $\alpha$ is one more unknown function of $t$ and $\bf x$.

Evaluating the commutator in the l.h.s. of (\ref{ic}) and equating coefficients for the linearly independent
differentials we obtain the following system of  equations for unknowns $\xi^0, \xi^a, \eta,\  V $ and $\alpha$:
\begin{gather}
\dot\xi^0=-\alpha, \quad  \xi^0_a=0,\la{de1}\\
\la{de2} \xi^b_{a}+\xi^a_{b}-\frac{2}n\delta_{ab}\xi^i_i=0,
\\\label{de6} \xi^i_i=-\frac{n}2\alpha,\\ \dot\xi^a+\eta_a
=0,\label{de7}\\
\label{de8}\xi^aV_a=\alpha V+\dot\eta+\ri[\eta,V]\end{gather}
where $\dot\eta=\frac{\p \eta}{\p t} $ and $\eta_a=\frac{\p \eta}{\p x_a}$.

Formally speaking, system of the determining  equations (\ref{de1})--(\ref{de8}) is rather complicated since it includes four arbitrary elements $V^0, V^1, V^2 $ and $V^3$ whose form should be fixed from the compatibility condition of this system. However, the major part of this system, i.e., equations (\ref{de1}), (\ref{de2}), (\ref{de6}) and (\ref{de7}), do not include these arbitrary elements. The immediate consequences of equation (\ref{de7}) are the following conditions:
\begin{gather}\la{con1}\frac{\p{\eta^a}}{\p x_b}=0\end{gather}
and
\begin{gather}\la{con2} \dot\xi^a+\eta^0_a=0.\end{gather}

Equation (\ref{de8}) in its turn is decoupled to the scalar and vector parts:
\begin{gather}\label{con3}\xi^aV^0_a=\alpha V^0+\dot\eta^0,\end{gather} and
\begin{gather}\label{con4}\xi^aV_a^b=\alpha V^b+\dot\eta^b-2\varepsilon^{bcd}\eta^cV^d\end{gather}
where $\varepsilon^{bcd}$ is the absolutely antisymmetric unit tensor.

In other words, system (\ref{de1})--(\ref{de8}) includes the autonomous subsystem  formed by equations (\ref{de1}), (\ref{de2}), (\ref{de6}), (\ref{con2}) and (\ref{con3}). Solving this subsystem we recover symmetries of the standard SE describing a spinless particle. Then, to find symmetries of the SP equation it is sufficient to solve equations (\ref{con2}) and (\ref{con4}) where $\xi^b$ and $\alpha$ are functions found at the previous step.

Thus the description of symmetries of the standard SE is the necessary step in the group classification of the SP equations.

\section{Symmetries of the standard Schr\"odinger equation}

In this section we specify inequivalent continuous symmetries of the standard SE with the following Hamiltonian:
 \begin{gather}\la{H3}H=-\frac12\p_a\p_a+V^0\end{gather}
 where $V^0=V^0({\bf x})$ is a scalar potential.
To achieve this goal it is sufficient to find inequivalent solutions of the determining equations (\ref{de1}), (\ref{de2}), (\ref{de6}), (\ref{con2}) and (\ref{con3}).
\subsection{Analysis of the determining equation}

It follows from (\ref{de1}) that both $\xi^0$ and $\alpha$ do not depend on $\bf x$. Equations (\ref{de2}) and (\ref{de6}) specify the dependence of coefficients $\xi^a$ on $\mathbf x$:
\begin{gather}\la{kil} \xi^a=
-\frac\alpha2 x_a+\theta^{ab} x_b+\nu_a\end{gather}
where $\alpha$, $\theta^{ab}=-\theta^{ba}$ and $\nu^a$ are arbitrary parameters.
 If $\alpha\neq0$ then, up to  shifts of spatial variables $x_a$ we can set $\nu^a=0$, and so equation (\ref{kil}) is decoupled to two versions: either
 \begin{gather}\la{kil1} \xi^a=
-\frac\alpha2 x_a+\theta^{ab} x_b\end{gather}
or
\begin{gather}\la{kil2} \xi^a=
\theta^{ab} x_b+\nu_a.\end{gather}

 Moreover, in accordance with (\ref{de7}), $\theta^{ab}$ are time independent, and so we have to specify the dependence on $t$ only for $\alpha$ or $\nu^a$.

Substituting (\ref{kil}) into   (\ref{con2}) and integrating the resultant equation we obtain the generic form of function $\eta^0$:
\begin{gather}\la{eta2}\eta^0=\frac{\dot\alpha}4x^2-\dot\nu_ax_a+f(t),
\ \ \alpha\nu^a=0\end{gather}
and so
the generic symmetry operator (\ref{so}) is reduced to the following form:
\begin{gather}\la{gso}Q=\xi^0 \p_t + \left(\frac{\dot\xi^0}2 x_a+\theta^{ab} x_b+\nu_a\right)\p_a-\frac{\ddot\xi^0}4x^2-\dot\nu_ax_a+f(t), \ \dot\xi^0\nu^a=0.\end{gather}
In addition, equation (\ref{de8}) takes one of the following   forms:
\begin{gather}\la{ded1}\left(\frac\alpha2x_a-\theta^{ab}x_b\right)V^0_a
+\alpha V^0+\frac{\ddot \alpha}4x^2-\dot f=0\end{gather} if
$\alpha\neq 0$, and
\begin{gather}\la{ded2}\left(\nu_a+\theta^{ab}x_b\right)V^0_a+\ddot\nu_ax_a+\dot {\tilde  f}=0\end{gather}
for $\alpha$ zero.
Thus to make the group classification of SEs  (\ref{H3}) are supposed to find  non-equivalent solutions of equations (\ref{ded1}) and (\ref{ded2}) for $V^0$ .

Let us specify the possible dependence of functions $\alpha, \eta$  and $\nu^0$ on $t$.
Differentiating equations (\ref{ded1}) and   (\ref{ded2}) w.r.t. $ x_c$  we obtain
\begin{gather}2\alpha (3V^0_c+x_b V^0_{cb})+\ddot\alpha x_c=\theta^{ab}x_bV^0_{ac}\la{ds2}\end{gather}
 for $\alpha$ nonzero, and
\begin{gather}\ddot\nu^c+\nu^bV_{cb}=-\theta^{ab}x_bV^0_{ac}.\la{ds1}\end{gather}
if $\alpha=0$.

In accordance with
 (\ref{ds2}) and (\ref{ds1})  and in view of the time independence of $V$ and $\theta^{ab}$ there are the following conditions for functions $\alpha$ and $\nu^a$
 \begin{gather}\la{anu1}a=0, \quad \ddot\nu=\mu \nu, \quad \text{if}\quad  V_{bc}=-\delta_{bc}\mu,\\
\la{anu2}a=0, \quad \ddot\nu=0, \quad \text{if}\quad  V_{bc} \neq-\delta_{bc}\mu,\\\la{anu3}\ddot\alpha=\mu\alpha\quad \text{if}\quad  V_{bc}=-\delta_{bc}\mu,\\\la{anu4}\ddot\alpha=0\quad \text{if}\quad  V_{bc} \neq-\delta_{bc}\mu.\end{gather}
where $\mu$ is a constant.

In other words, functions functions $\alpha$ and $\nu^a$  should be trigonometric, hyperbolic,   linear  or constant.

\subsection{Equivalence transformations}
By definition the equivalence transformations of the dependent and independent variables keep the generic form of equation (\ref{se}) but can change the potential $V=V({\mathbf x})$.
The obvious examples of such transformations is given by the following formulae:
\begin{gather}\la{et0}\begin{split}&{\bf x}\to \tilde {\bf x}={\bf x},\ \quad t\to\tilde t=t,\\&\psi(t,{\bf x})\to\tilde\psi(t,{\bf x})=\hat M\psi(t,{\bf x})\end{split}\end{gather}
where $\hat M$ is a constant non-degenerated matrix, and
\begin{gather}\la{et01}\begin{split}&{\bf x}\to \tilde {\bf x}={\bf x},\ \quad t\to\tilde t=t,\\&\psi(t,{\bf x})\to\tilde\psi(t,{\bf x})=\exp(\ri M t)\psi(t,{\bf x})\end{split}\end{gather}
where $M$ is a numeric matrix commuting with the potential. Up to equivalence transformations (\ref{quadqu}), this matrix is diagonal, i.e., $M=\mu+\nu\sigma_3$ with real parameters $\mu$ and $\nu$.

The equivalence transformations include all continuous (Lie) symmetries of  equation (\ref{se}) which do not change the potential $V$, and also transformations changing the potential. It is possible to show that for generic $V$ such (continuous) transformations include (\ref{et0}), (\ref{et01}), and transformations belonging   to the extended Euclid group $\tilde {\text E}$, i.e.,  shifts, rotations and scalings of independent variables see Section 5.

 In addition, for some particular potentials  there exist additional equivalence transformations. In the case of the trivial potential they have the following form:
\begin{gather}\la{et1}\begin{split}&{\bf x}\to \tilde {\bf x}=
\frac{\bf x}{\sqrt{1+t^2}}, \quad t\to\tilde
t=\frac1{\omega}{\arctan(
t)},\\&\psi(t,{\bf x})\to\tilde\psi(\tilde t,\tilde {\bf x})=(1+t^2)^{\frac{3}4}\text{e}^{\frac{-\ri\omega t{r}^{2}}{2(1+
t^2)}}\psi(t,{\bf x}),
\end{split}\end{gather}
\begin{gather}\la{et2}\begin{split}&{\bf x}\to \tilde {\bf x}=
\frac{\bf x}{\sqrt{1-t^2}}, \quad t\to\tilde
t=\frac1{\omega}{\text{arctanh}(
t)},\\&\psi(t,{\bf x})\to\tilde\psi(\tilde t,\tilde {\bf x})=(1-t^2)^{\frac{3}4}\text{e}^{\frac{\ri\omega t{r}^{2}}{2(1-
t^2)}}\psi(t,{\bf x})
\end{split}\end{gather}
and
\begin{gather}\la{et3}\begin{split}& x_a\to x'_a=x_a-\frac12{\kappa_a}t^2,\quad  t\to t'=t,\\& \psi(t,{\bf x})\to\psi'(t',{\bf x}')=\exp\left(-it\kappa_ax_a+\frac\ri3\kappa^2t^3\right)\psi(t,{\bf x}).\end{split}\end{gather}
where $\omega$ and $\kappa_a$ are arbitrary parameters, and $\kappa^2=\kappa_1^2+\kappa_2^2+\kappa_3^2$.

 Transformations  (\ref{et01}), (\ref{et1}), (\ref{et2}) and (\ref{et3}) keep the generic form of the related equation (\ref{se})  but change the trivial  potential  to
 \begin{gather}\la{quadqu}V=\mu+\nu\sigma_3,\\
\la{quad}V^0=\frac12\omega^2r^2,\\\la{qua}V^0=-\frac12\omega^2r^2,
\end{gather}
and
\begin{gather}\la{lin}V^0=\kappa_ax_a\end{gather}  correspondingly.

The equivalence of the isotropic  harmonic and repulsive oscillators to the free particle Schr\"odinger equation  was discovered in \cite{Nied2}. Formulae (\ref{et1}) and (\ref{et2}) present transformations for wave functions dependent on tree spatial variables while in \cite{Nied2} we can find them only for one dimensional case. For the equivalence transformations with arbitrary number of spatial variables see \cite{Nuca}.

Notice that   transformation (\ref{et1}) and (\ref{et2}) are valid  for any  equation (\ref{se}) with potential $V=V^0$ being a homogeneous function   of degree -2. In this case this equation is invariant with respect to dilatation transformations.

Mapping (\ref{et3}) connects the systems with trivial and free fall potentials \cite{Nied3}. But it is valid also for  potentials being functions of one or two spatial variables, say $ x_1$ or $x_1$ and $x_2$. In other words, it is valid provided the related equation (\ref{se})  is invariant w.r.t. shifts of independent variables along the third coordinate axis. In this case we have to set in (\ref{et3}) $a=2,3$ or $a=3$ correspondingly.

\subsection{Symmetries for SE  with trivial potentials}
Let us present the symmetries accepted by the Schr\"odinger equations with the trivial, isotropic oscillator and free fall  potentials. They are  known, but the related publications are not necessary easy accessible, and  we  fix them for the readers convenience. In addition, the specific combinations of just these symmetries are accepted by the other systems classified below, and   we need them to formulate the classification results.

Setting in  (\ref{ded1}) and   (\ref{ded2}) $V^0=0$ we easy solve the obtained equation and find the corresponding admissible symmetries (\ref{so}). They are linear combinations of the following symmetry operators:
\begin{gather}\begin{split}& P_a=-\ri\p_a,\quad M_{ab}=x_aP_b-x_bP_a,\\ &D=2tP_0-x_aP_a+\frac{3\ri }2, \end{split}\label{sok}\\\begin{split}
 &P_0=\ri\p_t,\quad G_a=tP_a-x_a,\\&
A=tD-t^2P_0-\frac{r^2}2.\end{split}\la{sos}\end{gather}

Let us remind that operators $P_0, P_a$ and $M_{ab}$ generate shifts and rotations of the independent variables and leave the wave function invariant. Operators $G_a$, $D$ and $A$ generate Galilei, dilatation and conformal  transformations correspondingly which act on dependent and independent variables. For the explicit form of these transformation see, e.g., \cite{FN}.

Operators (\ref{sok}), (\ref{sos}) together with the unit operator $I$ form the 13-dimensional Lie algebra sometimes called  Schr\"odinger algebra. Operators (\ref{sok}) and operator $I$ form a central extension of the Lie algebra of Schr\"odinger group.

      The additional identities satisfied by
 operators (\ref{sok}) and (\ref{sos}) are  \cite{FN}:
\begin{gather}\la{ide1}\begin{split}&P_aG_b-P_bG_a=M_{ab}, \\&P_aG_a+G_aP_a=2D+2t(P^2-2P_0),\\&
G_aG_a=2A+t^2(P^2-2P_0).\end{split}\end{gather}

On the set of solutions of equation (\ref{H}) the term in brackets is equal to $-2V^0$ which in our case is equal to zero, and so  relations (\ref{ide1}) express  generators $M_{ab},\ D$ and $A$ via bilinear combination of $P_a$ and $G_a$. Thus the invariance of the free Schr\"odinger equation with respect to  rotation, dilatation and conformal transformations appears to be a consequence of the symmetry with respect to the displacement and Galilei transformations.

Transformations (\ref{et2}) can be used to obtain symmetries of equation (\ref{H}) with the harmonic oscillator potential. They include $P_0$, $M_{ab}$ and the following generators:
\begin{gather}\la{sos2}\begin{split}&A^{+}=\sin(2\omega t)(P_0-{\omega^2}r^2)-\frac\omega2\cos(2\omega t)\left(x_aP_a+P_ax_a\right),\\
&\hat A^{+}=\cos(2\omega t)(P_0-{\omega}^2r^2)+\frac\omega2\sin(2\omega t)\left(x_aP_a+P_ax_a\right),\\
&B_{a}^+(\omega)=\sin(\omega t)P_a-\omega x_a\cos(\omega t),\quad \hat B_{a}^+(\omega)=\cos(\omega t)P_a+\omega x_a\sin(\omega t)\end{split}\end{gather}
where the upper marks "+" indicate the sign of the related potential  (\ref{quad}).

Symmetries for the repulsive oscillator potential (\ref{qua}) can be obtained from (\ref{sos2}) by the change $\omega\to \ri\omega$. As a result we obtain generators $P_0, \ M_{ab}$ in the same form as in (\ref{sok}), and the following operators:
\begin{gather}\la{sos3}\begin{split}& A^{-}=\exp(2\omega t)(P_0+\omega^2r^2-
\frac{\omega}2(x_aP_a+P_ax_a)),\\& \hat A^- =\exp(-2\omega t)(P_0+\omega^2r^2+
\frac{\omega}2(x_aP_a+P_ax_a)), \\&B^{-}_a=
\exp(\omega t)(P_a-\omega x_a),\ \hat B^{-}_a=
\exp(-\omega t)(P_a+\omega x_a) \end{split}\end{gather}
where the upper marks "-" indicate the sign of the related potential  (\ref{qua}).

Analogously, staring with  realization (\ref{sok}), (\ref{sos}) and making transformations (\ref{et3}) we find symmetries for equation (\ref{H}) with the free fall potential. To this effect it is sufficient to make the following changes:
\begin{gather}\la{chacha}\begin{split}&P_0\to P_0 +\kappa_aG_a+\frac12\kappa^2t^2,\quad P_a\to P_a+\kappa_at,\quad x_a\to x_a+\frac12\kappa_a t^2\end{split}\end{gather}
in all generators (\ref{sok}) and (\ref{sos}).

The presented symmetries  appear partly for the case of other particular potentials presented below. However, for generic potential the equivalence relations (\ref{et1})-(\ref{et3}) are not valid.

\subsection{Classification results for SE with arbitrary potential}

Consider equations (\ref{ded1}),  (\ref{ded2}) and their differential consequences (\ref{ds2})--(\ref{anu4}) for arbitrary potential $V$. Their solution is is a rather complicated procedure. We will use the algebraic approach which presupposes to use the basic property of symmetry operators:  that they should form a basis of a Lie algebra. This algebra by definition includes operator $P_0$ and the unit operator.

Using the mentioned  differential consequences  it is possible to show  that  the generic symmetry (\ref{so}) with coefficients (\ref{kil}), (\ref{eta2})  is a linear combination of symmetries (\ref{sok}), (\ref{sos}),  (\ref{sos2}), (\ref{sos3}) and yet indefinite function $f$. Thus to find all non-equivalent solutions of equation (\ref{ded1})  and (\ref{ded2}) we have to go over these combinations, restricting ourselves to the cases when they are non-equivalent.

Let one of conditions (\ref{anu2}) or (\ref{anu4}) is satisfied. In this case we come to a linear combinations of generators
$P_a,\ L_a=\frac12\varepsilon_{abc}$ and $D$  given by equation (\ref{sok}). They form a basis of the extended   Euclid  algebra $\tilde{\text{e}}$(3) whose non-equivalent subalgebras has been classified in \cite{baran}.  And just these subalgebras generate non-equivalent linear combinations of symmetries which we have to consider.

In particular, algebra $\tilde{\text{e}}$(3) has four non-equivalent one-dimensional subalgebras whose generators are \cite{baran}:
\begin{gather}\la{1d}L_3=M_{12},\quad L_3+P_3,\quad D+\mu L_3,\quad P_3.\end{gather}
The related parameters in (\ref{ded1}) and (\ref{ded2}) are
$\theta^{12}=1$ for $L_3$, $\theta^{12}=\nu^3=1 $ for $L_3+P_3$,
$\alpha=-2, \ \theta^{12}=\mu$ for $D+\mu L_3$ , $\nu^3=1$ for $P_3$,
and in all cases  $f=\kappa t$. In particular, for generator
$Q=L_3+\kappa t$ equation (\ref{ded1}) is reduced to the following
form:
\begin{gather*}L_3V=-\ri\kappa \end{gather*} and so
\begin{gather}\la{pot1}V=\kappa\varphi+G(\theta,r)\end{gather}
where $\varphi=\arctan\left(\frac{x_2}{x_1}\right)$ and
$\theta=\arctan(\frac{\tilde r}{x_3})$ are Euler angles.
 Just this solution is missing in the Boyer classification.

Solving the subclass of equations (\ref{ded1}) and (\ref{ded2})
corresponding to one-dimensional subalgebras specified in (\ref{1d})
we obtain the results  presented in Items 1--5 of Table 1. In Item 4
we set $\kappa=0$ since this parameter  can be reduced to zero using
mapping inverse to (\ref{et3}). Symmetry $G_3$  presented there
generates the same equation for potential as $P_3$ does.

It is possible to fix the following pairs   of "friendly symmetries"
 \begin{gather}\la{FS}\langle P_a, G_a\rangle, \quad \langle A, D\rangle\end{gather}
 which have the following property: any symmetry  induces the other
 symmetry from this pair. This phenomena is caused by the similarity of
 the  determining equations  corresponding to these symmetries.

  A more extended set of  "friendly symmetries" looks as follows:
 \begin{gather}\la{FS1}
  \langle (P_2,P_3), (G_2,G_3,L_1) \rangle,\quad \langle (G_2,G_3), (P_2,P_3,L_1) \rangle  \end{gather} and any pair from the first bracket induces the triplet from the second bracket.

The next step is to use the non-equivalent two-dimensional subalgebras of $\tilde{\text{e}}$(3) spanned on the following basis elements \cite{baran}:
\begin{gather}\la{2d} \langle L_3+\kappa t, P_3 \rangle,\quad \langle D+\kappa L_3, P_3 \rangle,\quad \langle P_2, P_3 \rangle,\quad \langle D, L_3 \rangle.\quad\end{gather}

 Any sets (\ref{2d}) includes at least one  element from (\ref{1d}). Thus we have to solve equations (\ref{ded1}) or (\ref{ded2}) generated by the second element for potentials presented in Items 1--5 of Table 1. As a result we obtain the results in Items 6--9. In fact the corresponding equations (\ref{se}) admit three (or even four) dimensional symmetry algebras since generators $P_3$ and $D$ are automatically attended by $G_3$ and $A$, see  equations (\ref{FS}), (\ref{FS1}) and the discussion nearby.

Analogously, considering the non-equivalent three dimensional subalgebras of $\tilde{\text{e}}$(3),whose basis elements are presented in the following formulae
\begin{gather}\la{3dim} \langle D, P_3, L_3\rangle,\quad  \langle D, P_1,  P_2\rangle, \quad \langle L_1, L_2, L_3\rangle,\quad \langle L_3, P_1,  P_2\rangle \\\langle P_1, P_2,  P_3\rangle,\quad \langle L_3+ P_3,\ P_1,\ P_2\rangle,\quad \langle D+\mu L_3, P_1,  P_2\rangle,\ \mu>0\la{empty}\end{gather}
we obtain potentials presented in Items 11--14. In the cases 12--14 we again have more extended symmetries thanks to the presence of the "friendly" elements.

In Table 1 as well as in the following Tables 2--4
$G(.)$ are arbitrary function of variables given in the brackets, $\mu, \ \kappa $ and $\omega_a$ are arbitrary real parameters, $\varepsilon_1,\ \varepsilon_2$ and $\varepsilon_3$ independently take values $\pm 1$ , subindexes $a$ and $k$ take all values 1, 2, 3 and 1, 2 correspondingly. In addition, we denote $r=\sqrt{x_1^2+x_2^2+x_3^2}, \ \tilde r=\sqrt{x_1^2+x_2^2}$ and $\varphi=\arctan\left({x_2}/{x_1}\right)$.

All presented systems by construction admit symmetries $P_0$ and $I$, the latter is  the unit operator. The additional symmetries are presented in Columns 3, where $P_a,\ L_a=\frac12\varepsilon_{abc}M^{bc}, \ D,\ A, B^\varepsilon_a(\omega_a)$ and $\hat B^\varepsilon_a(\omega_a)$ are generators  (\ref{sok}), (\ref{sos}), (\ref{sos2}) and (\ref{sos3}).
 The related symmetry algebras
are fixed in the fourth columns, where
 $\textsf{n}_{a,b}$ and $\textsf{s}_{a,b}$ are nilpotent and solvable Lie
 algebras correspondingly, whose  dimension is $a$ and the identification number is $b$.   To identify these algebras for $a\leq 6$ we use the notations proposed in \cite{snob}. The symbol $2\textsf{n}_{1,1}$ (or $3\textsf{n}_{1,1}$) denotes the direct sum of two (or three)  one-dimension algebras. In addition, g(1,2) and shcr(1,2) are Lie algebras of Galilei and Schr\"odinger groups in (1+2) dimensional space.


\begin{center}Table 1.
Non-equivalent  symmetries of  3$d$ Schr\"odinger equations whose potentials do not include quadratic terms.
\end{center}
\begin{tabular}{l l l l }
\hline No&Potential $V$&Symmetries&Invariance algebras
\\
\hline

1$\hspace{2mm} $&$G(\tilde r,x_3)+\kappa\varphi$
\vspace{1mm}
&$ L_3+\kappa t
$&\hspace{-2.5mm}$\begin{array}{c}\textsf{n}_{3,1}\ \ \text{if}\ \ \kappa\neq0,\\
3\textsf{n}_{1,1}\ \text{if}\ \ \kappa=0\end{array}$\\
2$\hspace{2mm} $&$G(\tilde r,x_3-
\varphi)+\kappa \varphi$
\vspace{1mm}
&$L_3+P_3+\kappa t$&$\hspace{-2.5mm}\begin{array}{c}\textsf{n}_{3,1}\ \ \text{if}\ \ \kappa\neq0,\\
3\textsf{n}_{1,1}\ \text{if}\ \ \kappa=0\end{array}$\\
3$\hspace{2mm} $&$\frac1{r^2} G(\frac{r}{\tilde r},r^\kappa e^
{-\varphi})$
 \vspace{1mm}
 &$D+\kappa L_3$&$\textsf{s}_{2,1}\oplus\textsf{n}_{1,1}$\\
 4$^\star\ $&$G(x_1,x_2)$
\vspace{1mm}
&$G_3,\ P_3$& $\textsf{n}_{4,1}$\\
5$^\ast\ $&$\frac1{ r^2}G(\varphi,\frac{\tilde r}r)$
\vspace{1mm}
&$A,\ D$&\textsf{sl}(2,R)$\oplus\textsf{n}_{1,1}$
\\
6$^\ast\ $&$\frac1{ r^2}G(\frac{\tilde r}r)$
\vspace{1mm}
&$A,\  D,\  L_3$&\textsf{sl}(2,R)$\oplus2\textsf{n}_{1,1}$\\
7$^\star\ $&$G(\tilde r)+\kappa\varphi$
\vspace{1mm}
&$L_3+\kappa t,\ G_3, \ P_3$&$\hspace{-2.5mm}\begin{array}{l}\textsf{s}_{5,14}\ \ \text{if}\ \ \kappa\neq0,\\\textsf{n}_{4,1}\oplus\textsf{n}_{1,1}\ \ \text{if}\ \ \kappa=0\end{array}$\\
8$\hspace{2mm} $&
\vspace{1mm}
$\frac1{\tilde
r^{2}}G(\tilde r^\kappa
e^{-\varphi})$
 \vspace{1mm}
&$D+\kappa L_3,\ G_3,\
P_3$ &$\textsf{s}_{5,38}$\\
9$^{\star\ast}$&\vspace{1mm}$\frac1{\tilde r^2}G(\varphi)$&$A,\ D,\   G_3,\  P_3$&$\textsf{s}_{6,242}$ \\
10$^\star $&$G(x_1)$
\vspace{1mm}
&$G_3,\ P_3,\  P_2,\ G_2,\ L_1$&\textsf{g}(1,2)\\
11$\hspace{2mm} $&
\vspace{1mm}
$G(r)$&$L_1,\  L_2,\  L_3$ &\textsf{so}(3)$\oplus  2\textsf{n}_{1,1}$\\
12$^\ast\ $&\vspace{1mm}$\frac\kappa {r^{2}}$& $A,\ D,\
L_1, \ L_2, \ L_3
$&\textsf{sl}(2,R)$\oplus\ $\textsf{so}(3)$
\oplus\textsf{n}_{1,1} $\\
13$^{\star\ast}$&\vspace{1mm}$\frac\kappa{\tilde r^2}$&$ A,\  D,\   G_3, \ P_3,\ L_3$&$\textsf{s}_{6,242} \oplus\textsf{n}_{1,1}$\\
\vspace{-2mm}
14$^{\star\ast}$&\vspace{2mm}$\frac\kappa{x_1^2}$&$ A,  D,  G_2, G_3, P_2,  P_3,  L_1 $&\textsf{schr}(1,2)\\

\hline
\hline
\end{tabular}

\vspace{2mm}

In the tables we specify also the admissible equivalence transformations additional to ones belonging to the extended Euclid group.  Namely, the star near the item number indicates  that the corresponding Schr\"odinger equation admits additional equivalence transformation (\ref{et3}) for independent variables $x_a$ provided $\frac{\p V}{\p x_a}=0$. The asterisk    marks the items which specify equations  admitting transformation (\ref{et1}) and (\ref{et2}).

In other words, up to the equivalence transformations (\ref{et1}) and (\ref{et2}) the potentials presented in items 5, 6, 9, 12, 13 and 14 can be generalized to include   the additional quadratic term $\frac{\varepsilon}2\omega^2r^2$. Analogously, the potentials presented in items 4, 7, 9, 10, 13,  and 14 can be transformed to the equivalent forms applying transformations (\ref{et3}). As a result the additional terms linear in $x_a$ will appear.

Thus we have described all symmetry algebras including generators
with time independent  coefficients $\xi^a$. In fact some of them
include the coefficients linear in $t$, but such generators appear
automatically, since they belong to "friendly symmetries". And just
these algebras are presented in Table 1.

The next step is to consider the versions corresponding to
symmetries dependent on time in a more complicated manner. To do it
we use  the versions presented in (\ref{anu1}) and (\ref{anu3}),
which correspond to symmetries of  the oscillator type presented by
formulae (\ref{sos2}) and (\ref{sos3}) and
  enumerate the possibilities with one, two, or three pairs of operators $B^\varepsilon_a, \hat B^\varepsilon_a$ with the same or different frequency parameters $\omega_a$. In other words, we again exploit the subalgebras of the extended Euclid algebra $\tilde{\text{e}}$(3), but the related basis elements are now given by equations (\ref{sos2}) and (\ref{sos3}).
   As a result we come to the list of inequivalent potentials and symmetries presented in Table 2. Five of them include arbitrary functions, but the related number of symmetries is rather restricted. The remaining five versions include arbitrary parameters, but the number of symmetries is more extended and equal to seven, nine or even eight.
   More exactly, the algebras of symmetries presented in Items 6-10 of
   Table 2 are solvable and have dimension $d=8$ and $d=9$ .
   We denote them formally as $\textsf{s}_{d,a}(.)$ without refereing to
   any data base,  since the classification such dimension   algebras
   is unknown.

 Let us present  commutation relations for basic elements of these algebras:
\begin{gather}\la{crr}\begin{split}&[P_0,B_a^\varepsilon]=\ri \omega\hat B_a^\varepsilon,\quad [P_0,\hat B_a^\varepsilon]=\ri \varepsilon \omega B_a^\varepsilon, \\& [P_0,G_3]=\ri P_3, \quad [B_a^\varepsilon,\hat B_b^\varepsilon]=\ri\delta_{ab} I,\\&[B_2^\varepsilon,L_3]=\ri B_1^\varepsilon,\quad [B_1^\varepsilon,L_3]=-\ri B_2^\varepsilon,\\& [\hat B_2^\varepsilon,L_3]=\ri \hat B_1^\varepsilon,\quad [\hat B_1^\varepsilon,L_3]=-\ri \hat B_2^\varepsilon\end{split}\end{gather}
where only non-trivial commutators are presented.

Thus, using the algebraic approach  we classify all non-equivalent
Lie symmetries admitted by 3$d$ Schr\"odinger equations. We recover
and complete the classical Boyer results but present them in more
convenient form with explicit specification of the admissible
symmetries and equivalence transformations. Moreover, some of the
presented results are new, see  Section 6.

\begin{center}Table 2.
Non-equivalent  symmetries of  3$d$ Schr\"odinger equations whose potentials  include quadratic terms.
\end{center}
\begin{tabular}{l l l l }
\hline No&Potential $V$&Symmetries&Invariance algebras
\\
\hline
1$\hspace{2mm} $&\vspace{2mm}$\varepsilon\frac{\omega^2x_3^2}2+G(x_1,x_2)$&
$B_3^\varepsilon(\omega),\ \hat B_3^\varepsilon(\omega)\ $&$\hspace{-2.5mm}\begin{array}{l}\textsf{s}_{4,6}\ \text{if}\  \varepsilon=-1,\\\textsf{s}_{4,7}\ \text{if}\  \varepsilon=1\end{array}$\\
2$\hspace{2mm} $&\vspace{2mm}$\varepsilon\frac{\omega^2x_3^2}2+G(\tilde r)+\mu\varphi$&$\begin{array}{l}B_3^\varepsilon(\omega),\ \hat B_3^\varepsilon(\omega),\\   L_3+\mu t\end{array}$&$\hspace{-2.5mm}\begin{array}{l}\textsf{s}_{5,15}\ \text{if}\ \ \varepsilon=-1, \mu\neq 0\\\textsf{s}_{5,16}\ \text{if}\  \varepsilon=1, \mu\neq 0\\\textsf{s}_{4,6}\oplus\textsf{n}_{1,1}\\ \text{if}\ \ \varepsilon=-1, \mu=0,\\\textsf{s}_{4,7}\oplus\textsf{n}_{1,1}\\ \text{if}\  \varepsilon=1, \mu=0\end{array}$\\
3$^\star\ $&\vspace{2mm}$\varepsilon\frac{\omega^2x_2^2}2+G(x_1)$&$\begin{array}{l}B_2^\varepsilon(\omega),\ \hat B_2^\varepsilon(\omega),\\   P_3,\ G_3\end{array}$&$\hspace{-2.5mm}\begin{array}{l}\textsf{s}_{6,160} \ \text{ if }\ \varepsilon=-1,\\\textsf{s}_{6,161} \ \text{ if }\ \varepsilon=1\end{array}$\\
4$\hspace{2mm} $&\vspace{1mm}$\varepsilon_1\frac{\omega_1^2x_1^2}2+
\varepsilon_2\frac{\omega_2^2x_2^2}2+G(x_3)$&$\begin{array}{l}B_1^{\varepsilon_1}(\omega_1), \hat B_1^{\varepsilon_1}(\omega_1),\\B_2^{\varepsilon_2}(\omega_2), \hat B_2^{\varepsilon_2}(\omega_2)\end{array} $&$\hspace{-2.5mm}\begin{array}{l}\textsf{s}_{6,162} \ \text{ if }\ \varepsilon_1=\varepsilon_2=-1,\\\textsf{s}_{6,164} \ \text{ if }\ \varepsilon_1\varepsilon_2=-1, \\\textsf{s}_{6,166} \ \text{ if }\ \varepsilon_1=\varepsilon_2=1\end{array}$\\
5$\hspace{2mm} $&\vspace{1mm}$\varepsilon\frac{\omega^2\tilde r^2}2+G(x_3)$&$\begin{array}{l}B_1^{\varepsilon_1}(\omega_1), \hat B_1^{\varepsilon_1}(\omega_1),\\B_2^{\varepsilon_2}(\omega_2), \hat B_2^{\varepsilon_2}(\omega_2), L_3\end{array} $&$\textsf{s}_{7,1}(\varepsilon_1,\varepsilon_2) $\\
6$\hspace{2mm} $&\vspace{1mm}$\varepsilon_1\frac{\omega_1^2x_1^2}2+\varepsilon_2\frac{\omega_2^2x_2^2}2+
\varepsilon_3\frac{\omega_3^2x_3^2}2$&$\begin{array}{l}B_a^{\varepsilon_a}(\omega_a), \hat B_a^{\varepsilon_a}(\omega_a),\\ a=1,2,3 \end{array} $&$\textsf{s}_{8,1}(\varepsilon_1,\varepsilon_2,\varepsilon_3) $\\
7$^\star\ $&\vspace{1mm}$\varepsilon_1\frac{\omega_1^2x_1^2}2+
\varepsilon_2\frac{\omega_2^2x_2^2}2$&$\begin{array}{l}B_1^{\varepsilon_1}(\omega_1), \hat B_1^{\varepsilon_1}(\omega_1), P_3,\\B_2^{\varepsilon_2}(\omega_2), \hat B_2^{\varepsilon_2}(\omega_2),  G_3\end{array}  $&$\textsf{s}_{8,2}(\varepsilon_1,\varepsilon_2)$\\
8$\hspace{2mm} $&\vspace{1mm}$\varepsilon\frac{\omega^2\tilde r^2}2+\varepsilon_3\frac{\omega_3^2x_3^2}2$&$\begin{array}{l} L_3,\ B_1^{\varepsilon}(\omega),\ \hat B_1^{\varepsilon}(\omega)\\  B_2^{\varepsilon}(\omega),\ \hat B_2^{\varepsilon}(\omega),\\   B_3^{\varepsilon_3}(\omega_3),\ \hat B_2^{\varepsilon_3}(\omega_3) \end{array} $&$\textsf{s}_{9,1}(\varepsilon,\varepsilon_3)$  \\
9$^\star\ $&\vspace{1mm}$\varepsilon\frac{\omega^2\tilde r^2}2$&$\begin{array}{l} G_3, P_3,   L_3, \ B_1^{\varepsilon}(\omega),\\ \hat B_1^{\varepsilon}(\omega),\ B_2^{\varepsilon}(\omega),\  \hat B_2^{\varepsilon}(\omega)\end{array}
$&$\textsf{s}_{9,2}(\varepsilon)$\\
10$^\star\ $&
\vspace{1mm}$\varepsilon\frac{\omega^2x_3^2}2$&$\begin{array}{l}B_3^\varepsilon(\omega), \hat B_3^\varepsilon(\omega),\ L_3,\ \\ P_1,\ P_2,\ G_1,\ G_2\end{array} $&$\textsf{s}_{9,3}(\varepsilon)$\\

\hline
\hline
\end{tabular}

\vspace{2mm}

\section{Symmetries of SP equations which do not include oscillator terms}
We have solved the subproblem of our classification problem, i.e., classified a reduced versions of SP equations with diagonal matrix potentials. In the present section we consider the general case with non-trivial external magnetic field.
In our notations it means that $V^a$ are not identically zero and the related  potential is a generic $2\times2$ matrix.

Since we have in hands all inequivalent solutions of equations (\ref{de2})-(\ref{de7}) which are found in the previous section, the only thing we need is to find the corresponding solutions of equations (\ref{con4}) where $\xi^a$ and $\alpha$ are known and, in accordance with (\ref{con1}),   $\eta^a$ are not dependent on $\mathbf x$ . In other words, it is necessary to consider all cases indicated in Table 1 and extend them to the case of non-trivial $V^a$ solving the corresponding equation (\ref{con4}) for potential components $V^a$.

First we consider the cases when SP equation admit one dimension algebras whose generators are presented in (\ref{1d}). The related scalar potentials are enumerated in items 1-4 of Table 1. In item 4 we can find one more symmetry, namely, $G_3$, but it can be treated as induced by $P_3$.

Let us start with the case which is not included into Table 1: no symmetry, all coefficients $\xi^a$ and $\eta^0$ are trivial. The corresponding equation (\ref{con4}) is reduced to the following form:
\begin{gather}\label{con5}\dot\eta^b-2\varepsilon^{bcd}\eta^cV^d=0\end{gather}

Up to constant matrix transformation the generic solution of (\ref{con5}) is:
\begin{gather}\la{CF}\begin{split}&V^1=V^2=0, \quad V^3=\lambda,\\&\eta^1=\cos(t),\quad \eta^2=\sin(t), \quad \eta^3=\rho,\quad
\text{and}\\&\eta^1=\sin(t),\quad \eta^2=-\cos(t), \quad \eta^3=\rho\end{split}\end{gather}
where $\lambda$ and $\rho$ are arbitrary constants.

Thus we find the symmetry of SP equation for a neutral particle interacting with the constant magnetic field which without loss of generality can be directed along the third coordinate axis.  However, like in the case of the harmonic oscillator,  the related SP equation (\ref{se}) can be reduced to the equation with trivial potential, which can be done using transformation (\ref{et0}).

Let us consider consequently the matrix extensions of all potentials and symmetries presented in Table 1. For the first  symmetry  specified in Item 4, i.e., for $P_3$ we have $\xi=1$ while $\xi^1, \ \xi^2$  and   $\alpha$ are trivial. Substituting these data into (\ref{con4}) we obtain:
\begin{gather}\la{n1}V^a_3=\dot\eta^a -2\varepsilon^{abc}\eta^bV^c.\end{gather}

Differentiating (\ref{n1}) w.r.t. $t$ we obtain the condition $\ddot \eta^a=2\varepsilon^{abc}\dot\eta^bV^c$, and so
\begin{gather}\la{n2}\dot \eta^a\ddot\eta^a=\frac12\frac{\p}{\p t}(\dot\eta^a\dot\eta^a)=0.
\end{gather}

In accordance with (\ref{n2}) vector components $\dot\eta^a$ should be time independent, and so
\begin{gather}\la{n3}\eta^a=k^at+n^a\end{gather} with some constants $k^a$ and $n^a$. Moreover, up to constant matrix transformations, $k^1=k^2=n^2=0$. As a result equation is reduced to the following system:
 \begin{gather}\la{n4} V^1_3=2n^3V^2,\quad  V^2_3=2(n^1V^3-n^3V^1)\\V^3_3=k^3-2n^1V^2,\la{n44} \quad k^3V^1=k^3V^2=0\end{gather}
 which has two classes of solutions defined up to constant matrix transformations:
 \begin{gather}\la{n5}\begin{split}&V^1=V^2=0,\quad V^3=k^3x_3+\Phi(x_1,x_2),\quad \eta^1=\eta^2=0,\quad \eta^3=k^3t+n^3,\\&V^3=\Phi(x_1,x_2),\quad V^1=V^2=0,\quad \eta^1=\eta^2=\eta^3=0,\end{split}\end{gather}
 and
 \begin{gather}\la{n6}\begin{split}&V^1=\Phi\cos(2nx_3)+\tilde\Phi(x_1,x_2) \sin(2nx_3),\\& V^2=\Phi(x_1,x_2)\sin(2nx_3)-\tilde\Phi(x_1,x_2) \cos(2nx^3), \quad V^3=\tilde G(x_1,x_2),\\& \eta^1=\eta^2=0,\quad \eta^3=n^3=n\end{split}\end{gather}
 where $\Phi(x_1,x_2),\ \tilde \Phi(x_1,x_2),\ G(x_1,x_2)$ and $\tilde G(x_1,x_2)$ are arbitrary functions of $x_1$ and $x_2$.

  Solutions (\ref{n5}) are not interesting since they correspond to the direct sum of ordinary SEs considered in the previous section. However, solutions (\ref{n6}) generate the matrix potential which cannot be diagonalized. Namely, these solutions generate
   the following potential (\ref{mp}):
 \begin{gather}\la{popo}V=N(x_1,x_2)+F(x_1,x_2)M(n,x_3)\end{gather}
 where $N(x_1,x_2), \ F(x_1,x_2)$ and $M(n,x_3)$ are matrices of the following generic form:
 \begin{gather}\la{popa}\begin{split}&N(x_1,x_2)=G(x_1,x_2)+\sigma_3\tilde G(x_1,x_2),\\&F(x_1,x_2)=\Phi(x_1,x_2)+\ri\sigma_3\tilde\Phi(x_1,x_2),\\
 &M(n,x_3)=\sigma_1\cos(2nx_3)+\sigma_2\sin(2nx_3)\end{split}\end{gather}
 with arbitrary functions $G, \tilde G, \Phi$ and $\tilde\Phi$.

  In the case of standard SE the symmetry $P_3$ induces the symmetry $G_3$. It can be verified by the direct calculation that for matrix potential (\ref{mp}), (\ref{n6}) it is not the case, and so the related SP equation is not Galilei invariant.

  Thus we have  generalized the potential $V=G(x_1,x_2)$ presented in Item 4 of Table 1 which admit symmetry $P_3$ to the case of matrix potential given by equation (\ref{popo}).However the SE with potential $V=G(x_1,x_2)$ admits equivalence transformation (\ref{et3}) while in the case of matrix potential (\ref{popa})
it looses this property. It means that the term $\kappa x_3$ which
we omit  in the case of scalar potential (since it can be reduced to
zero by the equivalence transformation  (\ref{et3})) now cannot be
omitted, and so we should add it to potential (\ref{popo}) as it is
done in Item 1 of Table 3. The same situation appears in all cases
when potentials include matrix $M(n,x_3)$ given by equation
(\ref{popa}).

 In complete analogy with the above we  solve equations (\ref{con4}) corresponding to the first symmetries  presented in Items 2 and 3, i.e., to $L_3$ and $L_3+P_3$. The only new element is the change of variables of $V^a$ from $x_3$ to  $\varphi$ and $\varphi+x_3$. As a result we obtain the following solutions
 \begin{gather}\la{n7}\begin{split}&V=N(\tilde r,x_3)+F(\tilde r,x_3)M(n,\varphi)+\kappa\varphi,\\&Q_1=L_3+\sigma_3n\end{split}\end{gather}
and
 \begin{gather}\la{n8}\begin{split}&V=N(\tilde r,x_3-\varphi)+F(\tilde r,x_3-\varphi)M(n,\varphi)
 ,\\&Q_2=L_3+P_3+\sigma_3n\end{split}\end{gather}
 where $Q_1$ and $Q_2$ are symmetries admitted by the SP equation with the given potentials.

 The meaning of symbols $N(.,.)$, $F(.,.)$ and $M(n,.)$ in equations (\ref{n7}) and (\ref{n8}) is the same as in equation (\ref{popo}). They denote matrices (\ref{popa}) depending on the arguments indicated in the brackets:
\begin{gather}\la{popal}\begin{split}&N(\tilde r,x_3)=G(\tilde r,x_3)+\sigma_3\tilde G(\tilde r,x_3),\\&F(\tilde r,x_3)=\Phi(\tilde r,x_3)+\ri\sigma_3\tilde\Phi(\tilde r,x_3),\\
 &M(n,\varphi)=\sigma_1\cos(2n\varphi)+\sigma_2\sin(2n\varphi),\end{split}\end{gather}
etc.

 A bit more efforts are requested for solution of equations (\ref{con4})
corresponding to the symmetry operator present in Item 4 of Table 1
since in this case parameter $\alpha$ is not trivial but equal to 2.
However up to this small generalization the solution procedure is
the same and gives the results presented in Item 4 of Table 3.

\newpage
\begin{center}Table 3.
Symmetries for  SP equation without quadratic potential.
\end{center}
\begin{tabular}{l l l l }
\hline No&Potential $V$&Symmetries&$\begin{array}{l}
\text{Algebras}\end{array}
$\\
\hline 1&$\begin{array}{l}N(x_1,x_2)+F(x_1,x_2)M(n,x_3)+n\kappa x_3
\end{array}$&$\begin{array}{l}P_3+\sigma_3n+n\kappa t\\\text{and }
G_3 \text{ if }n=0\end{array}$ \vspace{-1mm}
&$\begin{array}{c}\textsf{n}_{3,1}\ \ \text{if}\ \ n\kappa\neq0,\\
\vspace{-1mm}
3\textsf{n}_{1,1}\ \text{if}\  \kappa=0, n\neq0\\\vspace{-1mm}\textsf{n}_{4,1} \text{ if } n=0\end{array}$\\
2&$\begin{array}{l}N(\tilde r,x_3)+F(\tilde r,x_3)M(n,\varphi)+n\kappa\varphi\\
\end{array}$&$L_3+n\kappa t+\sigma_3n$
\vspace{-1mm}
&$\begin{array}{c}\textsf{n}_{3,1}\ \ \text{if}\ \ n\kappa\neq0,\\
3\textsf{n}_{1,1}\ \text{if}\ \kappa=0\end{array}$\\
3&$\begin{array}{l}N(\tilde r,x_3-\varphi)+
 F(\tilde r,x_3-\varphi)M(n,\varphi)\end{array}$&$L_3+P_3+\sigma_3n+n\kappa t$
\vspace{0mm}
&$\begin{array}{c} \textsf{n}_{3,1}\  \text{if}\
\kappa\neq0,\\3\textsf{n}_{1,1}\ \text{if}\  \kappa=0\end{array}$\\
4&$\begin{array}{l}\frac1{r^2} N(\theta,r^n\kappa e^
{-\varphi})+\frac1{r^2}F(\theta,r^n\kappa e^
{-\varphi})M(n,y),\\y=\varphi+\nu \ln(r),\ \nu+k\neq
0\end{array}$&$D+ k L_3+n(k+\nu)\sigma_3$ \vspace{0mm}
&\   $\textsf{s}_{2,1}\oplus\textsf{n}_{1,1}$\\
5&$\begin{array}{l}N(\tilde r)+n\kappa\varphi +F(\tilde
r)M(n,x_3)+\nu x_3,\\n\neq0
\end{array}$&$\begin{array}{l}L_3+n\kappa t,\\ P_3+n\sigma_3+\nu t\end{array}$
\vspace{0mm}
&$\begin{array}{l}\textsf{n}_{3,1}\oplus\textsf{n}_{1,1}\ \text{if
}    \kappa^2+\nu^2\neq0,\\
4\textsf{n}_{1,1}\ \text{if}\  \kappa=\nu=0\end{array}$\\
6 &$\begin{array}{l}N(\tilde r)+n\kappa\varphi +F(\tilde
r)M(n,\varphi)
\end{array}$&$\begin{array}{l}L_3+n\kappa t+n\sigma_3,\ P_3\\\text{and }
G_3 \text{ if }n=0\end{array}$ \vspace{-1mm}
&$\begin{array}{l}\textsf{n}_{3,1}\oplus\textsf{n}_{1,1}\ \text{if}
\
 n\kappa\neq0,\\
4\textsf{n}_{1,1}\ \text{if}\  \kappa=0,
n\neq0,\\\textsf{n}_{4,1}\oplus\textsf{n}_{1,1}\ \text{if}\ n=0\end{array}$\\
7&$\begin{array}{l}\frac1{\tilde r^2} N(\tilde r^n\kappa e^
{-\varphi})+\frac1{\tilde r^2}F(\tilde r^n\kappa e^
{-\varphi})M(n,\varphi)\end{array}$&$D+ k(L_3+n\sigma_3),\ P_3$
\vspace{-1mm}
&$\  \textsf{s}_{2,1}\oplus 2\textsf{n}_{1,1}$\\
8&$\begin{array}{l}\frac1{r^2}
N(\theta)+\frac1{r^2}F(\theta)M(n,\varphi),\ n\neq0
\end{array}$&$\begin{array}{l}D,
\ L_3+n\sigma_3\end{array}$ \vspace{-1mm} &$ \begin{array}{l}
\textsf{s}_{2,1}\oplus 2\textsf{n}_{1,1}
\end{array}$\\
9&$\begin{array}{l}\frac1{r^2}
N(\theta)+\frac1{r^2}F(\theta)M(n,\ln(r))
\end{array}$&$\begin{array}{l}D+n\sigma_3,
\ L_3\\\text{and } A \text{ if } n=0\end{array} $ \vspace{-1mm}
&$\begin{array}{l}\textsf{s}_{2,1}\oplus 2\textsf{n}_{1,1}\text{ if }n\neq0,\\
\textsf{sl}(2,R)\oplus2\textsf{n}_{1,1}\text{ if }n=0\end{array} $\\
 10&$\begin{array}{l} N(x_2)+F(x_2)M(n,x_3)+n\kappa
x_3\end{array}$&$\begin{array}{l}P_1,\ P_3+ n\sigma_3+n\kappa
t\\\text{and } G_1,\ G_3,\ L_2 \text{ if }n=0\end{array}$
\vspace{-1mm} &$\begin{array}{l} \textsf{n}_{3,1}\oplus
\textsf{n}_{1,1} \text{ if } n\kappa\neq0,\\ 4\textsf{n}_{1,1}
\text{ if } \kappa=0,\\\textsf{g}(1,2)\text{ if }n=0
\end{array}$\\
11 &$\begin{array}{l}G(r)+\Phi(r)\sigma_ax_a\end{array}$&
$L_a+\frac12\sigma_a,\ a=1,2,3
 $\vspace{-1mm}
&\ so(3)$
\oplus2\textsf{n}_{1,1}$\\
12 &$\begin{array}{l}\frac{\nu}{r^2}+
\sigma_3\frac{\mu}{r^2}+\frac{\alpha}{r^2}M(n,\ln(r))\end{array}$&$
\begin{array}{l}D+
n\sigma_3,\ L_1,\ L_2,\ L_3\\\text{and } A \text{ if
}n=0\end{array}$ \vspace{-1mm} &$\begin{array}{l}
\textsf{s}_{2,1}\oplus \textsf{n}_{1,1}\oplus \textsf{so}(3) \text{
if }n\neq0,\\\textsf{sl}(2,R)\oplus\textsf{so}(3)
\oplus\textsf{n}_{1,1}\\ \text{ if }n=0\end{array}$\\
13 &$\begin{array}{l}\frac{\nu}{\tilde r^2}+
\sigma_3\frac{\mu}{\tilde r^2}+\frac{\alpha}{\tilde
r^2}M(n,\varphi),\ n\neq0\end{array}$&$D,\  L_3+
n\sigma_3, P_3$&$\ \textsf{s}_{3,1}\oplus2\textsf{n}_{1,1}$\\
14 &$\begin{array}{l}\frac{\nu}{\tilde r^2}+
\sigma_3\frac{\mu}{\tilde r^2}+\frac{\alpha}{\tilde
r^2}M(n,\ln(\tilde r))
\end{array}$&$\begin{array}{l}D+n\sigma_3,\
 P_3,\ L_3\\\text{and } A \text{ if
}n=0\end{array}$ \vspace{-1mm} &$\begin{array}{l}
\textsf{s}_{3,1}\oplus2\textsf{n}_{1,1}\text{ if }n\neq0,
\\\textsf{sl}(2,R)
\oplus3\textsf{n}_{1,1}\text{ if }n=0\end{array}$\\
15 &$\begin{array}{l}\frac{\nu}{x_3^2}+
\sigma_3\frac{\mu}{x_3^2}+\frac{\alpha}{x_3^2}M(n,\ln(x_3))\end{array}$&$
\begin{array}{l}D+n\sigma_3,\
 P_1,\ P_2, L_3\\\text{and } A,\ G_1,\ G_2 \text{ if
}n=0\end{array}$ \vspace{-1mm} &$\begin{array}{l}
\textsf{s}_{5,43}\oplus\textsf{n}_{1,1}\text{ if }n\neq0,
\\\textsf{schr}(1,2)\text{ if }n=0\end{array}$\\
16
&$\begin{array}{l}\frac{\nu}{r^2}+\frac{\mu}{r^3}\sigma_ax_a\end{array}$&
$D, A, L_a+\frac12\sigma_a,\ a=1,2,3
 $\vspace{-1mm}
&$\ \textsf{s}_{2,1}\oplus\textsf{n}_{1,1}\oplus$so(3)\\
\hline
\hline
\end{tabular}


In Table 3 the symbols $M(.), N(.), F(.)$ denote matrices defined in
equations (\ref{popa}) and (\ref{popal}) where $n$ can take
arbitrary values including zero, $G(r)$ and $\Phi(r)$ are arbitrary
functions.

The  potentials presented in Items 1--4 of Table 3 include arbitrary
functions, and the related SP equations admit one dimensional
symmetry algebras generated by operators (\ref{1d}). For some
particular functions $G, \tilde G, \Phi$ and $\tilde \Phi$ these
symmetries can be extended to two dimensional algebras presented by
equation (\ref{2d}). To fix these particular functions we can start
with the first of two symmetries presented in some brackets given in
(\ref{2d}) and use the corresponding potential presented in one of
the items 1-4. Then, asking for existence of the second symmetry  we
specify the mentioned arbitrary functions. The other way which in
general leads to another result is to start with the second symmetry
with the related potential and ask for existence of the first
symmetry. In this way we  find the versions presented in Items 5-9
of Table 3.

 Analogously, asking for extensions of the two dimension symmetry algebras
 we obtain the potentials and the related generators presented in Items
10-16. The potential in Item 10 includes two arbitrary functions
while the remaining potentials are defined up to arbitrary
parameters.

 Thus we have generalized all symmetries and potentials proportional to the unit matrix presented in Table 1, to the case of generic matrix potentials. The related classification results are summarized in Table 3.

\section{Symmetries of SP equations with  oscillator terms}
 The final step of our classification consists in the generalization of the scalar potentials which include the harmonic oscillator terms. Such potentials are presented in  Table 2. In addition, we are supposed to analyze all versions presented in Table 1 and marked by the asterisk. The related potentials can be generalized to include the isotropic harmonic oscillator term. Moreover, such term cannot be removed using the equivalence transformations (\ref{et1}) and (\ref{et2}) provided the potential includes non-diagonal matrix terms.

 Starting with the analogous reasons, it is necessary to generalize the potentials including linear terms, which can be generated using equivalence transformations (\ref{et3}) starting with the data of Table 1 marked by the star.

 Let us consider consequently all symmetries presented in Table 2 and find matrix potentials compatible with them. To do it we are supposed solve equations (\ref{con4}) for $V^a$ where functions $\xi^a$ can be found comparing definition (\ref{so}) and explicit expressions for symmetry operators presented in the Table.

Considering symmetries presented in Item 1 of Table 2 we find that neither $B^\varepsilon_3$ nor $\hat B^\varepsilon_3$ generate non-trivial solutions of equation (\ref{con4})  for components $V^a$ of matrix potential (\ref{mp}). However, the linear combinations $Q^{\pm}=B^-_3\pm\hat B^-_3$ are compatible with non-trivial $V^a$. Indeed, in this case the only nonzero components $\xi^a$ and $\eta^0$ are
\begin{gather}\la{my1}\xi^3=\exp(\pm \omega t),\quad \eta^0=\mp\exp(\pm \omega t)\omega x_3.\end{gather}
The related components $\eta^a$ should have the same dependence on $t$ as $\xi^3$, and so in analogy with  (\ref{n5}) we have to set
\begin{gather}\la{my2}\eta_1=\eta_2=0,\ \eta^3=n\exp(\pm \omega t).\end{gather}

In the following we omit signs $\pm$  but reserve the possibilities for parameter $\omega$ be positive or negative, and write the corresponding symmetry as
\begin{gather}\la{my4} Q=\exp(\omega t)(P_3-\omega x_3+n\sigma_3).\end{gather}

Substituting these data into equation (\ref{con4}) we come  to the following system:
\begin{gather}\la{my3} V^1_3=2n V^2,\quad  V^2_3=-2n V^1,\\\la{my33} V^3_3=\omega n.\end{gather}

Equation (\ref{my33}) is easy solved. System (\ref{my3}) coincides with (\ref{n4}) where $k^1=k^2=k^3=n^1=n^2=0$. Thus its solution is given by formula (\ref{n6}).  Just this solution is presented in Item 1 of Table 4 together with the scalar term $-\frac12\omega^2 x_3^2$. Moreover, the potential presented here generalizes the scalar potential presented in Item 1 of Table 2 to the matrix case.

Thus we have found an example of matrix potential which include the repulsive oscillator term such that equation SP equation (\ref{se}) admits a one parametric Lie group additional to shifts of the time variable.  This potential includes three arbitrary matrices of special form fixed in (\ref{popa}) and dependent on two or one spatial variables.  To classify the potentials which admit symmetry (\ref{con4}) and are compatible with more extended symmetry groups we should apply the additional conditions (\ref{con5}) where $\xi^a$ and $\eta^a$ are functions specifying the additional symmetries (\ref{so}), (\ref{eta1}).  Since all admissible  inequivalent symmetries are presented in Table 2, we can easily fix these functions  and then solve the related equations (\ref{con5}).

Let the SP equation with generic potential presented in Item 1 of Table 4 admits the additional symmetry $L_3+\kappa t$ fixed in Item 2 of Table 2.  The corresponding functions $N(x_1,x_2)$ and $F(x_1,x_2)$ should be rotationally invariant, i.e., to depend on $\tilde r=\sqrt{x_1^2+x_2^2}$, and we come to the potential presented in Item 2 of Table 4. For the case of additional symmetry $P_1$ these functions can depend only on $x_2$, and we come to the potential fixed in Item 3 of Table 4, and so on and so on.

In this way we find all potentials presented in Items 1 - 7 of Table 4. Notice that this list does not include matrix extensions of scalar potentials presented in Items 5, 8 and 10 of Table 2 since the related conditions  (\ref{con5}) appears to be incompatible.

Consider now the case when the potential includes the 3d oscillator term $V^0=\frac\varepsilon2 \omega^2r^2$. This term cannot be reduced to zero provided the other components $V^a$ are functions of $\bf x$. Moreover, it is nothing but a particular case of the potential presented in Item 6 of Table 2, whose possible extensions to the case of matrix potential have been already classified in the above, see Items 3--7 of Table 3. The only specific point is that now we have to consider the additional symmetries $A^\pm$ and $\tilde A^\pm$ whose explicit form is given in (\ref{sos2}) and (\ref{sos3}).

It is easy to show that neither $A^+, \tilde A^+$ nor their linear combinations are compatible with equations (\ref{con5}). However, the linear combinations $\tilde Q^\pm=A^-\pm\tilde A^-$ are the only symmetries including $A^+$ and $ \tilde A^+$ which solve this equation with the corresponding $V^a$. These symmetries can be represented as:
\begin{gather}\la{my6}\tilde Q=\exp(2\omega t)\left({P_0}+{\omega}x_aP_a-\frac{3\omega\ri}2+\omega^2 r^2\right)\end{gather}
where we omit the signs $\pm$ but reserve the possibilities for parameter $\omega$ to be positive or negative. Comparing (\ref{my6}) with (\ref{so}) we find the corresponding functions $\xi^a$ and $\eta^0$ in the following form:
\begin{gather}\la{my7}\xi^a=\exp(2\omega t)\omega x_a, \ \eta^0=\exp(2\omega t)\left(\omega^2 r^2-\frac{3\omega\ri}2\right)\end{gather}

The corresponding  functions  $\eta^a$ should have the same dependence on $t$ as $\xi^a$, and so up to matrix transformations and  in analogy with  (\ref{n5}) we can  set
\begin{gather}\la{my8}
\eta^3=\exp(2\omega t)\omega n\sigma_3, \ \eta^1=\eta^2=0\end{gather}

Substituting (\ref{my8})  into equation (15) we obtain the following  system:
\begin{gather}\la{my9}\begin{split}& x_aV^1_a=-2 V^1+2n V^2,\quad  x_a V^2_a=-2 V^2-2n V^1,\quad x_aV^3_a=-2 V^3+2n\end{split}\end{gather}
whose generic solution is presented in Item 8 of Table 4. This solution is compatible with the only symmetry $\hat Q=\tilde Q+\eta_3$ additional to $P_0$.

In analogy with the above we can specify arbitrary function
$N(\varphi,\theta)$ and $F(\varphi,\theta)$ in such way that
symmetries of the corresponding SP equation became more extended.
All such inequivalent specifications are enumerated in Items 9--13
of Table 4.
\begin{center}Table 4. Potentials with oscillator
terms and  symmetries for  Schr\"odinger-Pauli equation.
\end{center}
\begin{tabular}{l l l l }
\hline No&Potential
$V$&Symmetries&$\begin{array}{l}\text{Algebras}\end{array}
$\\
\hline \ 1&$\begin{array}{l}N(x_1,x_2)+\sigma_3\omega n
x_3-\frac12\omega^2x_3^2+F(x_1,x_2)M(n,x_3)\end{array}$&$\begin{array}{l}Q\end{array}$&
$\textsf{s}_{2,1}\oplus\textsf{n}_{1,1}$\\
\ 2&$\begin{array}{l}N(\tilde r)+\sigma_3\omega n x_3 +F(\tilde
r)M(n,x_3)+n\kappa \varphi -\frac12\omega^2 x_3^2
\end{array}$&$\begin{array}{l}Q,\ L_3+n\kappa t\ \ \end{array}$
\vspace{1mm}
&$\
3\textsf{n}_{1,1}$\\
\ 3&$\begin{array}{l}N(x_2)+\sigma_3\omega n x_3-\frac12\omega^2 x_3^2+F(x_2)M(n,x_3)\end{array}$&$\begin{array}{l}Q, \ P_1,\ G_1\end{array}$&$\textsf{s}_{2,1}\oplus2\textsf{n}_{1,1}$\\
\ 4&$\begin{array}{l}N(x_1)+F(x_1)M(n,x_3)
+\frac{\varepsilon}2\omega_2^2 x_2^2-\frac{1}2\omega_3^2
x_3^2+n\omega\sigma_3x_3
\end{array}$&$\begin{array}{l}Q, \ B_2^{\varepsilon},\ {\tilde B}_2^{\varepsilon}\end{array}$&$\textsf{s}_{5,17}$\\
\ 5&$\begin{array}{l}\mu M(n,x_3) +\frac{\varepsilon}2\omega_2^2
x_2^2-\frac{1}2\omega_3^2 x_3^2+\sigma_3n\omega x_3
\end{array}$&$\begin{array}{l}Q, \ B _2^{\varepsilon},\ {\tilde B}_2^{\varepsilon},\ P_1\end{array}$&$\textsf{s}_{5,17}\oplus\textsf{n}_{1,1}$\\
\ 6&$\begin{array}{l}\mu M(n, x _3) +\frac{\varepsilon_1}2\omega_1^2
x_1^2+ \frac{\varepsilon_2}2\omega_2^2 x_2^2-\frac{1}2\omega_3^2
x_3^2+\sigma_3 n\omega x_3
\end{array}$&$\begin{array}{l}Q, \ B_1^{\varepsilon_1},\ {\tilde B}_1^{\varepsilon_1},\\\ B_2^{\varepsilon_2},\ {\tilde B}_2^{\varepsilon_2}\end{array}$&$\textsf{s}_{7,3}$\\
\ 7&$\begin{array}{l}\mu M(n,
 x_3) +\frac{\varepsilon_1}2\omega^2 \tilde r^2-\frac{1}2\omega_3^2 x_3^2+\sigma_3 n\omega x_3
\end{array}$&$\begin{array}{l}Q,  \ B_1^{\varepsilon},\
{\tilde B}_1^{\varepsilon},\\ B_2^{\varepsilon},\
{\tilde B}_2^{\varepsilon},\ L_3\end{array}$&$\textsf{s}_{7,3}\oplus\textsf{n}_{1,1}$\\

\vspace{1mm}

\ 8&$\begin{array}{l}\sigma_3\omega n -\frac{1}2\omega^2  r^2+
\frac{1}{r^2}N(\varphi,\theta) +\frac{1}{r^2}F(\varphi,\theta)
M(n,\ln(r))
\end{array}$&$ \begin{array}{l}\tilde Q \end{array}$&$\textsf{s}_{2,1}\oplus\textsf{n}_{1,1}$\\
\ 9&$\begin{array}{l}\sigma_3\omega n -\frac{1}2\omega^2  r^2
+\frac{1}{r^2}N(\theta)+\frac{1}{r^2}F(\theta) M(n,{\ln(r)})
\end{array}$&$\begin{array}{l} \tilde Q, \ L_3 \end{array}$&$\textsf{s}_{2,1}\oplus2\textsf{n}_{1,1}$\\
10&$\begin{array}{l}\sigma_3\omega n -\frac{1}2\omega^2  r^2
+\frac{1}{\tilde r^2}N(\varphi)+\frac{1}{\tilde r^2}F(\varphi)
M(n,\ln(\tilde r))
\end{array}$&$\begin{array}{l} \tilde Q,\ B^-_3, \tilde B^-_3\end{array}$
&\\
11&$\begin{array}{l}\sigma_3\omega n -\frac{1}2\omega^2  r^2
+\frac{\mu}{r^2}+\frac\lambda{r^2} M(n,\ln(r))
\end{array}$&$\begin{array}{l}\tilde Q, \ L_1 ,\ L_2, \ L_3\end{array} $&so(3)$\oplus\textsf{s}_{2,1}\oplus\textsf{n}_{1,1}$\\
12&$\begin{array}{l}\sigma_3\omega n -\frac{1}2\omega^2  r^2
+\frac{\mu}{\tilde r^2}+\frac\lambda{\tilde r^2} M(n,\ln(\tilde r))
\end{array}$&$\begin{array}{l}\tilde Q, \ L_3, \ B^-_3, \tilde B^-_3\end{array} $&\\
13&$\begin{array}{l}\sigma_3\omega n -\frac{1}2\omega^2  r^2
+\frac{\mu}{x_3^2}+\frac{\lambda}{x_3^2} M(n,\ln(x_3))
\end{array}$&$\begin{array}{l}\tilde Q,\  \ L_3, \ B^-_1,\\ \tilde B^-_1,\ B^-_2, \tilde B^-_2\end{array} $&\\
14&$\begin{array}{l}N(\tilde r)+F(\tilde r)M(n,\varphi)
+\frac{\varepsilon}2\omega^2 x_3^2
\end{array}$&$\begin{array}{l}L_3+\sigma_3n,\ B_3^\varepsilon,\ \tilde B_3^\varepsilon\end{array}$&$\textsf{s}_{4,6}\oplus\textsf{n}_{1,1}$\\
15&$\begin{array}{l}N(x_2)+\frac{\varepsilon}2\omega^2
x_1^2+F(x_2)M(n,x_3)+
n\kappa x_3 \end{array}$&$\begin{array}{l}P_3+n\sigma_3+n\kappa t, \\  \ B^\varepsilon_1,\ \tilde B^\varepsilon_1\end{array}$&$\textsf{s}_{4,6}\oplus\textsf{n}_{1,1}$\\
16&$\begin{array}{l}\mu M(n,x_3) +\frac{\varepsilon_1}2\omega_1^2
x_1^2+\frac{\varepsilon_2}2\omega_2^2 x_2^2+n\kappa x_3
\end{array}$&$\begin{array}{l}B^{\varepsilon_1}_1,
\tilde B^{\varepsilon_1}_1, B^{\varepsilon_2}_2, \tilde
B^{\varepsilon_2}_2,\\  P_3+n \sigma_3+n\kappa
t\end{array}$&$\hspace{-2.5mm}\begin{array}{l} \textsf{s}_{7,1}
\text{ if } \kappa\neq0,\\ \textsf{s}_{6,n}
\oplus\textsf{n}_{1,1}\text{ if }\kappa=0\end{array}$\\
17&$\begin{array}{l}\mu M(n,x_3) +\frac{\varepsilon}2\omega^2 \tilde
r^2 + n\kappa x_3
\end{array}$&$\begin{array}{l} B^\varepsilon_1,\  \tilde B^\varepsilon_1,
\ B^\varepsilon_2,\ \tilde B^\varepsilon_2, \\ L_3,\ P_3+n
\sigma_3+n\kappa t \end{array}$&$\begin{array}{l}\textsf{s}_{7,2}
\oplus\textsf{n}_{1,1}\text{ if } \kappa=0,\\ \textsf{s}_{8,1}\text{ if } \kappa\neq0\end{array}$\\

\hline
\hline
\end{tabular}


The remaining items 14--17 of  Table 4 present the  potentials which
also include the oscillator or repulsive oscillator terms, but the
corresponding symmetries do not include either $Q$ or $\tilde Q$.
Notice that these potentials are analytical in $\omega$ and so they
are well defined for trivial   $\omega$ also.

In the table  $Q$ and $\tilde Q$ are generators fixed in (\ref{my4})
and (\ref{my6}), algebras $\textsf{s}_{6,n}$ are specified in Item 4
of Table 2 . The symbols $N(\cdot), F(\cdot), M(n,\cdot)$ are used
to denote matrices (\ref{popal}) depending on the arguments fixed in
the brackets, $\mu$ and $\lambda$ are arbitrary real parameters.

Symmetries specified in Table 4 correspond to  non-zero values of
parameter $n$. However, the corresponding equations are well defined
also for $n=0$. In this case  symmetry algebras  are more extended,
namely, symmetries $Q$ and $\tilde Q$ are replaced by the pairs $<
B^-_3, \tilde B^-_3>$ and $< A^-,\ \hat A^->$ correspondingly. In
addition, symmetry $G_3$ appears in Items 15, 16, 17 and symmetry
$G_1$ should be included into Item 5.
\section{Discussion}

The main  goal of the present paper was to give the group classification of SP equations for neutral particles. This program has been realized, the classification results are summarized in Tables 3 and 4. In accordance with these results there exist 33 inequivalent equations of this type which admit different symmetry groups. The most extended groups are eight parametrical and are presented in Items 7 and 13 of Table 4.

The group classification  presents a priory information about all admissible symmetries of the considered class of equations and the explicit form of the corresponding arbitrary elements which in our case are matrix potentials. Such information is useful for construction of physical models with requested symmetries.  Moreover, it is an important and in fact the necessary step in search for systems with different kinds of generalized symmetries, in particular, superintegrable systems, since it supplies us by important equivalence relations.

A particular and important  case of matrix potentials is the case of diagonal matrices, when we have a direct sum of standard Shcr\"odinger equations with position dependent potential. The group classification of such SEs is a subproblem of our more general problem, and we present its solutions in Tables 1 and 2. In contrast with the well known presentation in paper \cite{Boy}   our classification results are completed and include four case missing in \cite{Boy}, see Items 1, 2, 7 of Table 1 and Item 2 of Table 2.  In addition, we make our best to specify clearly the equivalence relations which are different for different potentials.

The potential terms $\varphi=\kappa\arctan\left({x_2}/{x_1}\right)$ missing in the Boyer classification belongs to the class of harmonic potentials which find many interesting applications including such exotic ones as the robot navigation.

One more new feature of our presentation is the clear specification of the invariance algebras  using notations proposed in \cite{snob}. We believe that this information is  important and useful. In particular,  the reader interested in the Casimir operators of the symmetry algebras can easy find them in book \cite{snob}.

Notice that the low dimension algebras of dimension $d\leq5$ and
some class of the algebras of dimension 6 had been classified
by
 Mubarakzianov \cite{mur}, see also  more contemporary
 and accessible papers \cite{bas} and \cite{boy2} were his results are slightly  corrected.

 The presented list of symmetries  does not include the infinite symmetry group of transformations $\psi\to\psi+\tilde\psi$ where $\tilde\psi$ is an arbitrary solution of equation (\ref{H}). In accordance with the superposition rule,  such evident symmetries are valid for all linear equations.

 Let us note that in general our matrix potentials cannot be interpreted as a sum of scalar potentials and Pauli terms dependent on a purely magnetic field. Indeed, the found potentials $V^a$ are not supposed to be divergent less. However,many of these potentials are compatible with the condition
$\p_a V^a=0$ which can be added additionally. Without  this condition the found potentials represent more generic external fields.

Symmetries of SE with matrix potential have some specific features in comparison with the case of a scalar potential. which will be fixed in the following comments.

 It is well known that the standard SE is invariant w.r.t.Galilei transformation of space variable $x_a$ iff its potential does not depend on this variable or is equivalent to such potential. For example, it is the case for the oscillator potential since the corresponding SE is equivalent to the free one, and the generators of Galilei transforms have exotic form (\ref{sos3}).

For the case of matrix potentials this observation does not pay. Indeed, potentials fixed in Items 1--7 of Table 4 are not either trivial or equivalent to trivial. Nevertheless, they admit symmetries being linear combinations of the exotic generators of Galilei transformations.

In contrast with the standard SE for which exotic conformal generators $A^-$ and $\hat A^-$ (\ref{sos3}) are valid only for the repulsive oscillator potential, there exist such matrix potentials which are not equivalent to linear ones and admit linear combinations of such generators. Just these potentials are presented in Items 8--13 of Table 4.

Let us note that the list potentials and symmetries presented in Table 4 can be generalized to the case of imaginary parameters $\omega$. In this case the repulsive oscillator term is transformed to the harmonic oscillator one. However, the symmetry operators became non-hermitian with respect to the standard scalar product used in quantum mechanics. This paradox in principle can be overcame  in frames of quantum mechanics with indefinite metrics, for example,  CP-symmetric quantum mechanics \cite{bender}.

In this paper we restrict ourselves to searching for symmetries of SP equations for neutral particles. A more generic problem of group classification of SP equations for charged particles can to be a subject of our following work.  For group classification of nonlinear Schr\"odinger equations and their conditional symmetries see papers \cite{pop,Nn1} and \cite{FN1}.




\begin{thebibliography}{99}
\bibitem{Nied}
U. Niederer, The maximal kinematical invariance group of the free
Schr\"odinger equations,
 Helv. Phys. Acta, \textbf{45} 802--810 (1972).


\bibitem{And}
R. L. Anderson, S. Kumei, C. E. Wulfman,  Invariants of the
equations of wave mechanics. I., Rev. Mex. Fis. { \textbf 21},  1--33
(1972).


\bibitem{Boy}
C. P. Boyer, The maximal kinematical invariance group for an
arbitrary potential, Helv. Phys. Acta { \textbf 47}, 450--605 (1974).

\bibitem{Wint}P. Winternitz,  J. Smorodinsky, M. Uhlir. and I. Fris,  Symmetry groups in classical and quantum
mechanics, Sov. J. Nucl. Phys. { \textbf 4}, 444--450 (1967).

\bibitem{Wint1} A. A. Makarov, J. A. Smorodinsky, K. Valiev and P.Winternitz, A systematic search for nonrelativistic systems with dynamical symmetries. Il Nuovo Cimento A  { \textbf 52}, 1061-1084  (1967).
\bibitem{Mil} W. Miller, Jr.. Symmetry and separation of variables. Addison-Wesley P.C. (1977).

\bibitem{evan} N. W. Evans, Superintegrability of the Winternitz system,
 Phys. Lett. { \textbf A147}, 483-486 (1990).


\bibitem{evan2}N. W. Evans, Superintegrability in classical mechanics,
 Phys. Rev. { \textbf A41},  5666-5676 (1990).


\bibitem{PW} W. Miller, Jr., S. Post and P. Winternitz ,  Classical and Quantum Superintegrability with
Applications,  J. Phys. A: Math. Theor. { \textbf 46}, 423001 (2013).

\bibitem{NNN} Anatoly G. Nikitin, Supersymmetries in Schr\"odinger
-- Pauli Equations and in Schr\"odinger
     Equations with Position Dependent Mass.  Supersymmetry and Coherent States,
     pp. 133-162. Springer, Cham (2019).

     \bibitem{wintmar}Marquette, Ian, and Pavel Winternitz. Higher Order Quantum Superintegrability: A New “Painlevé Conjecture”. Integrability, Supersymmetry and Coherent States. Springer, Cham, pp. 103-131. Springer, Cham (2019).

 \bibitem{NikNik}A. G. Nikitin, Higher-order symmetry operators for
 Schr\"odinger equation. In
      CRM Proceedings and Lecture Notes (AMS),  \textbf { 37} , pp. 137--144
       (2004).

\bibitem{11} C. Quesne and V. M. Tkachuk, Deformed algebras,
position-dependent effective masses and curved spaces: an exactly
solvable Coulomb problem, J. of Phys. A: Math. and Gen. \textbf{ 37},
4267 (2004).


\bibitem{rac}C. Quesne, Quadratic algebra approach to an exactly solvable
position-dependent mass Schr\"odinger equation in two dimensions,
SIGMA \textbf{ 3}, 067 (2007).

\bibitem{Koch} R. Ko\c{c} and M. Koca,
A systematic study on the exact solution of the position dependent
mass Schr\"odinger equation, J. Phys. A \textbf{ 36}, 8105 (2003).

\bibitem{Cru}Sara Cruz, Y. Cruz and Rosas-Ortiz Oscar, Dynamical Equations,
Invariants and Spectrum Generating Algebras of Mechanical Systems
with Position-Dependent Mass, SIGMA \textbf{ 9}, 004 (2013).

\bibitem{Fordy} Allan P. Fordy and Qing Huang, Generalised Darboux-Koenigs
Metrics and 3-Dimensional Superintegrable Systems, SIGMA {\textbf 15}, 037
(2019).

\bibitem{Fordy2}A. P. Fordy and Qing Huang, Superintegrable systems
on 3 dimensional flat space, J. of Mathematical Physics, Analysis,
Geometry {\bf 153}, 103687 (2020).

\bibitem{NZ} A. G. Nikitin and T. M. Zasadko, Superintegrable systems with position dependent mass.
     Journal of Mathematical Physics {\bf 56}, 042101 (2015).

    \bibitem{NZ2} A. G. Nikitin and T. M. Zasadko,Group classification of Schrodinger
     equations with position dependent mass. J. Phys. A: Math. Theor.
     \textbf{49}, 365204  (2016).

\bibitem{NN} A.G. Nikitin,  Kinematical invariance groups of the 3d {S}chr\"{o}dinger
 equations with position dependent masses,  J.~Math. Phys. \textbf{ 58},
 083508 (2017).

   \bibitem{N1}  A. G. Nikitin. Superintegrable and shape invariant systems with position dependent mass,
     J. Phys. A: Math. Theor.  \textbf{48}, 335201 (2015).

 \bibitem{N3}     A. G. Nikitin, Algebras of discrete symmetries and
     supersymmetries for the Schroedinger-Pauli equation,
     Int. J.  Mod. Phys.  \textbf{A 14},
     885-897 (1999).

\bibitem{N4} J. Niederle and A. G. Nikitin, Extended supersymmetries for
     the Schroedinger-Pauli equation,  J.  Math.
     Phys.  \textbf{40},  1280-1293 (1999).

\bibitem{kar}A. G. Nikitin  and Y. Karadzhov,   Enhanced
classification of matrix superpotentials, J. Phys. A: \textbf{44},
445202 (2011).

\bibitem{N5} A. G. Nikitin, Matrix superpotentials and superintegrable systems
     for arbitrary spin, J. Phys. A: Math. Theor.  \textbf{45}, 225205
     (2012).


\bibitem{N6} A. G. Nikitin, Superintegrability and supersymmetry of Schr\"odinger-Pauli equations for neutral
     particles, J. Math. Phys. (2012) \textbf{53}, 122103 (2012).


\bibitem{N7} A. G. Nikitin, Superintegrable systems with spin invariant with
     respect to the rotation group. J. Phys. A: Math. Theor. \textbf{46},
     256204 (2013).

     \bibitem{N8} A. G. Nikitin, New exactly solvable systems with Fock symmetry,
     J. Phys. A: Math. Theor. \textbf{45},  485204 (2012).


\bibitem{N9} A. G. Nikitin. Laplace-Runge-Lenz vector for arbitrary spin,
     J. Math. Phys. \textbf{54}, 123506 (2013).

\bibitem{N10} A. G. Nikitin. Superintegrable systems with arbitrary spin. Ukr. J. Phys. \textbf{58},
     1046-1054 (2013).

      \bibitem{B}  J. Beckers, N. Debergh, and A. G. Nikitin, On
     Parasupersymmetries and  Relativistic Description for Spin One
     Particles: II. The interacting context with (electro)magnetic
     fields, Fortschritte der Physik \textbf{43},  81-95
     (1995).

\bibitem{M}E. Ferraro, N. Messina and A.G. Nikitin,
     Exactly solvable relativistic model with the anomalous
     interaction, Phys. Rev. A \textbf{81}, 042108 (2010).


\bibitem{Nuca} A. G. Nikitin, The maximal'kinematical'invariance group for an arbitrary potential revised
     J. of Mathematical Physics, Analysis, Geometry \textbf{14}, 519-531
      (2018).




\bibitem{olver}P. Olver. Application of Lie Groups to Differential
Equations. Springer-Verlag, New York (2000);
electronic version: PJ Olver-2000-books.google.com.




\bibitem{Nied2}U. Niederer, The maximal kinematical invariance group of
the harmonic oscillator, Helv. Phys. Acta {\bf 47}, 191--200 (1973).

\bibitem{Nied3} U. Niederer. The group theoretical equivalence of the free particle, the harmonic oscillator
 and the free fall.  Proceedings of the 2nd International Colloquium on Group Theoretical Methods in
Physics, University of Nijmegen, The Netherlands (1973).


     \bibitem{FN} W.I. Fushchich and A. G. Nikitin, Symmetries of Equations of
     Quantum Mechanics, N.Y., Allerton Press Inc. (1994).

\bibitem{baran}W. I. Fushchich, L. F. Barannyk  and A. F. Barannyk,
 Subgroup analysis of Galilei and Poincare groups and reduction of
 nonlinear equations (in Russian), Naukova Dumka, Kiev (1991).
 Extended English version of this monograps which includes subalgebras of
 algebra $\tilde{\text{e}}$(3) is unpublished.

\bibitem{snob}Libor \^Snobl and Pavel Winternitz.  {Classification and identification of Lie algebras.} CRM
Monograph Series, v. 33 (2010).




\bibitem{mur} G. M. Murakzianov, Classification of real structures of Lie algebras of fifth order, Izvestia Vysshykh Uchebnykh Zavedenii. Matematika  {\textbf 3},  99-106 (1963).

\bibitem{bas}P. Basarab-Horwath, L. Lahno and R. Zhdanov, The structure of the Lie algebras and the cxlassification problem  of partial differential equations, Acta Applicandae Methematica {\textbf 69}, 43-94 (2001).



\bibitem{boy2} V. Boyko, J. Patera and R. Popovych, R., Computation of invariants of Lie algebras
by means of moving frames, Phys. A: Math. Gen. \textbf{39}, 5749-5762,  (2006) .

\bibitem{bender} C. M. Bender. PT symmetry In quantum and classical physics. World Scientific Publishing (2018).

\bibitem{pop} A. G. Nikitin and R. O. Popovych,
Group classification of nonlinear Schr\"odinger equations,  Ukr.
Math. J. {\bf 53},  1255-1265 (2001).

\bibitem{Nn1} A.G. Nikitin,  Group classification of systems of non-linear
     reaction-diffusion equations with general diffusion matrix. I.
     Generalized Ginsburg-Landau equations, J. Math. Analysis and Applications
     (JMAA) {\textbf 324},  615-628 (2006).

\bibitem{FN1} W.I. Fushchich and  A.G. Nikitin, Higher symmetries and exact
     solutions of linear and nonlinear Schr\"odinger equation
     {J.  Math. Phys.} {\textbf 38}, 5944-- 5959 (1997).

\end{thebibliography}
\end{document}